\documentclass{llncs}
\usepackage{amssymb}
\usepackage{amsmath}
\usepackage{graphicx}

\newcommand{\Perm}[1]{\mathrm{\bf S}_{#1}}
\newcommand{\set}[1]{\left\{#1\right\}}
\def\F{\mathbb{F}}
\newcommand{\esymm}[2]
{
\sigma_{#1}\left(#2\right)
}
\begin{document}
\title{Cellular Automata with Symmetric Local Rules}
\titlerunning{Automata with Symmetric Local Rules}
\author{Vladimir V. Kornyak}
\institute{Laboratory of Information Technologies \\
           Joint Institute for Nuclear Research \\
           141980 Dubna, Russia \\
           \email{kornyak@jinr.ru}}
\authorrunning{Vladimir V. Kornyak}
\maketitle
\begin{abstract}
The cellular automata with local permutation invariance
are considered. We show that in the
two-state case the set of such automata coincides with the generalized
Game of Life family.
We count the number of equivalence classes of the rules under consideration with respect to permutations of states.
This reduced number of rules can be efficiently generated
in many practical cases by our C program.
Since a cellular automaton is a combination of a local rule and a lattice, we consider also maximally symmetric two-dimensional lattices.
In addition, we present the results of compatibility analysis
of several rules from the Life family.
\end{abstract}

\section{Introduction}
The number of possible local rules for $q$-state cellular automaton
defined on a $(k+1)$-cell neighborhood is double exponential of $k$, namely,  $q^{q^{k+1}}.$
It is natural to restrict our attention to special classes of local
rules.~ S. Wolfram showed \cite{Wolfram} that even simplest 2-state 3-cell automata, which he terms \emph{elementary cellular automata}
(there are $2^{2^3}=256$
possible rules for these automata),
demonstrate the unexpectedly complex behavior.

We consider here a class of \emph{symmetric} local rules
defined on a $(k+1)$-cell neighborhood. Here by a `symmetry' we mean
a symmetry with respect to
all permutations of $k$ cells (\emph{points}
or \emph{vertices}) surrounding $(k+1)$th cell, which time evolution
the local rule determines. The reasons to distinguish such
rules
are
\begin{itemize}
    \item The number of possible symmetric rules is
 the single  exponential of polynomial of $k$
 (see formula (\ref{nqk}) below). For example, for $k=8$, $q=2$
      (the case of the Conway's game of Life)
      the number of symmetric rules is $262144\approx2.6\times10^5$,
      whereas the number of all possible rules
$\approx1.3\times10^{154}.$
    \item The symmetry of the neighborhood under permutations is
        in a certain sense a discrete analog of general local diffeomorphism
        invariance which is believed must hold for any fundamental physical
        theory based on continuum spacetime.%
\footnote{
    The symmetric group of any finite or infinite set $M$ is often
    denoted by $\mathrm{Sym}(M)$. If $M$ is a manifold, then a
    diffeomorphism of $M$
     is nothing but a special --- continuous and differentiable ---
     permutation from $\mathrm{Sym}(M)$.}

    \item This class of rules contains such widely known
     automata as
      \emph{\textbf{Conway's Life}}. In fact,
     as we show below, any symmetric rule is a natural
      generalization of the  \emph{\textbf{Life}} rule.
\end{itemize}

\section{Symmetric Local Rules and Generalized Life}
\subsection{Symmetric Rules}
We interpret a $(k+1)$-cell neighborhood of a cellular automaton
as a \emph{$k$-star} graph, i.e., rooted tree of height 1 with $k$ leaves. We call this the
\emph{$k$-valent} neighborhood. We adopt the convention that
the leaves are indexed by the numbers $1,2,\ldots,k$ and the root is
numbered by $k+1$. For example, the trivalent neighborhood looks like
this
\begin{center}
\setlength{\unitlength}{1.5pt}
\begin{picture}(40,40)(0,0)
\put(20,20){\circle{4}}
\put(28,20){\makebox(0,0){\footnotesize{$x_4$}}}
\put(7,11.5){\circle*{4}}
\put(14.5,10){\makebox(0,0){\footnotesize{$x_1$}}}
\put(33,11.5){\circle*{4}}
\put(40.5,10){\makebox(0,0){\footnotesize{$x_2$}}}
\put(20,34){\circle*{4}}
\put(28,34){\makebox(0,0){\footnotesize{$x_3$}}}
\put(18.5,18.5){\line(-5,-3){10}}
\put(21.5,18.5){\line(5,-3){10}}
\put(20,22){\line(0,1){10}}
\end{picture}
\end{center}
A \emph{local rule} is a function
specifying one time step evolution of the state of root
\begin{equation}
x^{\prime}_{k+1} = f\left(x_1,\ldots,x_k,x_{k+1}\right).
    \label{localrule}
\end{equation}
We consider the set $R_{\Perm{k}}$ of local rules symmetric with respect to the group $\Perm{k}$ of all
permutations of leaves, i.e., variables $x_1,\ldots,x_k.$
We will consider also  the subset $R_{\Perm{k+1}}\subset R_{\Perm{k}}$ of rules symmetric with respect to permutations of all $k+1$ points
of the neigborhood. For brevity we shall use the terms \emph{$k$-symmetry }and \emph{$(k+1)$-symmetry}, respectively.
\par
Obviously the total numbers of $k$- and $(k+1)$-symmetric rules are, respectively,
\begin{eqnarray}
        N^q_{\Perm{k}}~~&=&~~q^{\binom{k+q-1}{q-1}q},
        \label{nqk}\\
        N^q_{\Perm{k+1}}&=&~~q^{\binom{k+q}{q-1}}.
        \label{nqk+1}
\end{eqnarray}
\subsection{Life Family}
The ``Life family'' is a set of 2-dimensional, binary cellular automata
similar to\emph{\textbf{ Conway's Life}} \cite{GofL}, which rule is defined on 9-cell (3$\times$3) Moore neighborhood and is described as follows. A
cell is ``born'' if it has exactly 3 alive neighbors, ``survives''
if it has 2 or 3 such neighbors, and dies otherwise. This rule is
symbolized in terms of the ``birth''/``survival'' lists as
B3/S23. Another examples of automata from this family are
\emph{\textbf{HighLife}} (the rule B36/S23), and
\emph{\textbf{Day\&Night}} (the rule B3678/S34678).
 The site \cite{lives-site} contains collection of more than twenty rules from the Life family with Java applet to run these rules and
 descriptions of their behavior.
\par
Generalizing this type of local rules, we define a \emph{$k$-valent Life
rule} as a \emph{binary} rule on a $k$-valent neighborhood, described by
two \emph{arbitrary} subsets of the set $\set{0,1,\ldots,k}.$ These subsets $B,S\subseteq\set{0,1,\ldots,k}$ contain conditions for the $x_k\rightarrow x'_k$ transitions of the forms $0\rightarrow1$ and
$1\rightarrow1$, respectively. Since the number of subsets of any finite
set $A$ is $2^{\left|A\right|}$, the number of rules defined by two sets
$B$ and $S$ is equal to $2^{k+1}\times2^{k+1}= 2^{2k+2}$, which in turn is
equal to (\ref{nqk}) evaluated at $q=2.$ On the other hand, \emph{different} pairs $B$/$S$
define \emph{different} rules.
\par
Thus, we have the obvious\\
\textbf{Proposition.} \emph{For any $k$ the set of $k$-symmetric binary rules coincides
with the set of $k$-valent Life rules.}
\par
This proposition implies, in particular, that one can always express any
 symmetric binary rule in terms of ``birth''/``survival'' lists.
\subsection{Equivalence with Respect to Permutations of States}
Exploiting the symmetry with respect to renaming of $q$ states
of cellular automata allows us to reduce the number of rules to consider.
Namely, it suffices to consider only orbits (equivalence classes) of the rules under $q!$ permutations
forming the group $\Perm{q}$. For counting orbits of a finite group $G$
acting on a set $R$ (\emph{rules}, in our context) there is the formula called \emph{Burnside's lemma}.
This lemma states
(see., e.g., \cite{Harary}) that
the number of orbits, denoted $\left|R/G\right|$, is equal to the average number of points
$R^g\subset R$ fixed by elements $g\in G$:
\begin{equation}
\left|R/G\right| = \frac{1}{\left|G\right|}\sum_{g\in G}\left|R^g\right|.
    \label{Burnside}
\end{equation}
Thus, the problem is reduced to finding the sets of fixed points.
\par
Since we are mainly interested here in the binary
automata, we shall consider further the case $q=2$ only. For this case
the combinatorics is rather simple.
\par
Specializing (\ref{nqk}) and (\ref{nqk+1}) for  $q=2$ we have
the numbers of binary $k$- and $(k+1)-$symmetric rules, respectively,
\begin{eqnarray}
        N_{\Perm{k}}&=&2^{2k+2},
        \label{nk}\\
        N_{\Perm{k+1}}&=&2^{k+2}.
        \label{nk+1}
\end{eqnarray}

The group $\Perm{2}$ contains two elements $e=(0)(1)$ and $c=(01)$ (in cyclic notation). After S.Wolfram, we shall call the permutation $c \in \Perm{2}$ \emph{``black-white''} (shortly \emph{BW}) transformation.
\par
Since the number of fixed points for $e$ is either (\ref{nk}) or (\ref{nk+1}), all we need is to count the number of fixed points for the
BW transformation in both $k$- and
$(k+1)$-symmetry cases.
The rules in both these cases can be represented,
respectively, by the bit strings
\begin{equation}
    \alpha_1\alpha_2\cdots\alpha_{2k+2}
    \label{bitstrk}
\end{equation}
and
\begin{equation}
    \alpha_1\alpha_2\cdots\alpha_{k+2}
    \label{bitstrk1}
\end{equation}
in accordance with the tables
\begin{center}
\begin{tabular}[t]{ccc}
\begin{tabular}[t]{ccccc|l}
$x_1$&$x_2$&$\cdots$&$x_{k}$&$x_{k+1}$&$x'_{k+1}$
\\
\hline
0&0&$\cdots$&0&0&$\alpha_1$\\
0&0&$\cdots$&0&1&$\alpha_2$\\
1&0&$\cdots$&0&0&$\alpha_3$\\
1&0&$\cdots$&0&1&$\alpha_4$\\
$\vdots$&$\vdots$&$\vdots$&$\vdots$&$\vdots$&$\vdots$\\
1&1&$\cdots$&1&0&$\alpha_{2k+1}$\\
1&1&$\cdots$&1&1&$\alpha_{2k+2}$
\end{tabular}
&
\hspace*{50pt}
&
\begin{tabular}[t]{ccccc|l}
$x_1$&$x_2$&$\cdots$&$x_{k}$&$x_{k+1}$&$x'_{k+1}$
\\
\hline
0&0&$\cdots$&0&0&$\alpha_1$\\
1&0&$\cdots$&0&0&$\alpha_2$\\
1&1&$\cdots$&0&0&$\alpha_3$\\
$\vdots$&$\vdots$&$\vdots$&$\vdots$&$\vdots$&$\vdots$\\
1&1&$\cdots$&1&0&$\alpha_{k+1}$\\
1&1&$\cdots$&1&1&$\alpha_{k+2}$
\end{tabular}
\end{tabular}
\end{center}
One can see from these tables that the BW transformation acts similarly
on both strings (\ref{bitstrk})  and (\ref{bitstrk1}), namely,
$$
    \alpha_1\alpha_2\cdots\alpha_{n-1}\alpha_{n}
    \stackrel{~~BW~~}{\longrightarrow}
    \bar{\alpha}_n\bar{\alpha}_{n-1}\cdots\bar{\alpha}_2
    \bar{\alpha}_{1},
$$
where bar means complementary transformation of bit, i.e.,
$\bar{\alpha}=\alpha+1\mod 2$, and $n=2k+2$ or $n=k+2$.
\par
The fixed point condition
\begin{equation}
    \alpha_1\alpha_2\cdots\alpha_{n-1}\alpha_{n}
        =
        \bar{\alpha}_n\bar{\alpha}_{n-1}\cdots\bar{\alpha}_2
        \bar{\alpha}_{1}
        \label{fixedcond}
\end{equation}
implies that the string is defined by a half of bits, i.e., by $k+1$ or by
$(k+2)/2$ bits. In other words, the numbers of different bit strings satisfying
condition (\ref{fixedcond}) are, respectively, $2^{k+1}$ and
$2^{(k+2)/2}$ (for $k$ even). For the $(k+1)$-symmetry case with odd $k=2m+1$ condition
(\ref{fixedcond}) leads to the contradiction
$$
\bar{\alpha}_{m+1}=\alpha_{m+1},
$$
which means zero number of bit strings in this case.
\par
Summarizing our calculations (recollecting that $G=\Perm{2}$
in formula (\ref{Burnside}), i.e.,  $\left|G\right|=2$), we have:
\begin{itemize}
    \item in the case of $k$-symmetry
\begin{itemize}
    \item number of BW-symmetric rules
\begin{equation}
    N_{\Perm{k}BW}=2^{k+1},
    \label{skbw}
\end{equation}
    \item number of non-equivalent rules
\begin{equation}
    N_{\Perm{k}/BW}=2^{2k+1}+2^k,
    \label{sk/bw}
\end{equation}
\end{itemize}
    \item in the case of $(k+1)$-symmetry
\begin{itemize}
    \item number of BW-symmetric rules
\begin{equation}
    N_{\Perm{k+1}BW}
    =\left\{
    \begin{array}{lll}
    2^{k/2+1}&\text{if}&k=2m,\\
    0&\text{if}&k=2m+1,
    \end{array}
    \right.
    \label{sk+1bw}
\end{equation}
    \item number of non-equivalent rules
\begin{equation}
    N_{\Perm{k+1}/BW}
    =\left\{
    \begin{array}{lll}
    2^{k+1}+2^{k/2}&\text{if}&k=2m,\\
    2^{k+1}&\text{if}&k=2m+1.
    \end{array}
    \right.
    \label{sk+1/bw}
\end{equation}
\end{itemize}
\end{itemize}
For example, for trivalent rules the numbers are
$$
N_{\Perm{3}BW}=16,~~~
N_{\Perm{3}/BW}=136,~~~
N_{\Perm{3+1}BW}=0,~~~
N_{\Perm{3+1}/BW}=16.
$$

\section{Assembling Neighborhoods into Regular Lattices}
$k$-valent neighborhoods can be gathered into a lattice in many ways.
In fact, any $k$-regular graph may serve as a space for a cellular automaton. In applications a cellular automaton acts, as a rule, on a lattice embedded in a metric space, usually 2- or 3-dimensional Euclidean
space. Any graph, being 1-dimensional simplicial complex, can be embedded into $\mathbb{E}^3$. When dealing with symmetric local rules it is natural
to consider equidistant regular systems of cells. We consider here only
two-dimensional case as more simple and suitable for visualization of
the automaton behavior. Most symmetric 2D lattices correspond to the tilings
by congruent regular polygons. The number of different types of such tilings is rather restricted (see, e.g., \cite{GrunbaumSheppard}).
To denote a $k$-valent lattice composed of regular $p$-gons we use
the \emph{Schl\"afli symbol} $\{p,k\}$.
\subsection{2D Euclidean Metric}
\label{2D Euclidean Metric}
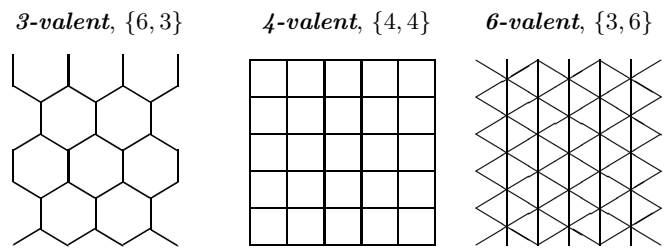
\begin{figure}
\centering
\setlength{\unitlength}{1.2pt}
\begin{picture}(210,75)(0,0)
\put(27,70){\makebox(0,0){\emph{\textbf{3-valent}}, $\left\{6,3\right\}$}}
\multiput(0,50)(17.32,0){4}{\line(0,1){10}}
\multiput(0,20)(17.32,0){4}{\line(0,1){10}}
\multiput(0,20)(17.32,0){3}{\line(5,-3){8.6}}
\multiput(17.32,20)(17.32,0){3}{\line(-5,-3){8.6}}
\multiput(0,30)(17.32,0){3}{\line(5,3){8.6}}
\multiput(8.66,5)(17.32,0){3}{\line(0,1){10}}
\multiput(8.66,35)(17.32,0){3}{\line(0,1){10}}
\multiput(8.66,35)(17.32,0){3}{\line(5,-3){8.6}}
\multiput(8.66,5)(17.32,0){3}{\line(5,-3){8.6}}
\multiput(0,0)(17.32,0){3}{\line(5,3){8.6}}
\multiput(8.66,45)(17.32,0){3}{\line(5,3){8.6}}
\multiput(8.66,45)(17.32,0){3}{\line(-5,3){8.6}}
\put(105,70){\makebox(0,0){\emph{\textbf{4-valent}}, $\left\{4,4\right\}$}}
{\setlength{\unitlength}{1.4pt}
\multiput(64,0)(10,0){6}{\line(0,1){50}}
\multiput(64,0)(0,10){6}{\line(1,0){50}}
}
\put(175,70){\makebox(0,0){\emph{\textbf{6-valent}}, $\left\{3,6\right\}$}}
{\setlength{\unitlength}{1.4pt}
\multiput(125,0)(0,10){3}{\line(5,3){50}}
\multiput(125,50)(0,-10){3}{\line(5,-3){50}}
\put(125,20){\line(5,-3){33.5}}
\put(125,30){\line(5,3){33.5}}
\put(125,10){\line(5,-3){16.7}}
\put(125,40){\line(5,3){16.7}}
\put(141.72,0){\line(5,3){33.5}}
\put(158.44,0){\line(5,3){16.7}}
\put(141.72,50){\line(5,-3){33.5}}
\put(158.44,50){\line(5,-3){16.7}}
\multiput(133.36,0)(8.36,0){5}{\line(0,1){50}}
}
\end{picture}
\caption{All regular lattices in $\mathbb{E}^2$}
    \label{lattices-e2}
\end{figure}

There are only three regular lattices in $\mathbb{E}^2$
(see Fig. \ref{lattices-e2}).
\par
Since real computers have finite memory, cellular automata can be simulated
only on a finite lattice. Usually the universe of a cellular automaton
is a rectangle instead of an infinite plane. There are different ways
to handle the edges of the rectangle. One possible method is to fix states of the border cells. This breaks the symmetry of the lattice and is thus
not interesting for us.
Another way is to glue together the opposite edges of the rectangle.
\par
The toroidal arrangement is a standard practice, but
it would be interesting to study cellular automata on nonorientable
surfaces also. There are 3 different identifications of opposite sides of
a rectangle: the \emph{torus} $\mathbb{T}^2$, the \emph{Klein bottle}
$\mathbb{K}^2$, and the \emph{projective plane} $\mathbb{P}^2$.
All these spaces (the gluing does not affect their Euclidean metric) are shown in the figure below.
\par
\begin{center}
\setlength{\unitlength}{1pt}
\begin{picture}(250,50)(0,0)
\put(10,40){\makebox(0,0){\underline{$\mathbb{T}^2$:}}}
\multiput(45,10)(0,30){2}{\makebox(0,0){A}}
\multiput(65,5)(0,40){2}{\vector(-1,0){40}}
\multiput(30,25)(31,0){2}{\makebox(0,0){B}}
\multiput(25,5)(40,0){2}{\vector(0,1){40}}
\put(100,40){\makebox(0,0){\underline{$\mathbb{K}^2$:}}}
\multiput(135,10)(0,30){2}{\makebox(0,0){A}}
\put(155,5){\vector(-1,0){40}}
\put(115,45){\vector(1,0){40}}
\multiput(120,25)(31,0){2}{\makebox(0,0){B}}
\multiput(115,5)(40,0){2}{\vector(0,1){40}}
\put(190,40){\makebox(0,0){\underline{$\mathbb{P}^2$:}}}
\multiput(225,10)(0,30){2}{\makebox(0,0){A}}
\put(245,5){\vector(-1,0){40}}
\put(205,45){\vector(1,0){40}}
\multiput(210,25)(31,0){2}{\makebox(0,0){B}}
\put(205,5){\vector(0,1){40}}
\put(245,45){\vector(0,-1){40}}
\end{picture}
\end{center}
We need to check whether the regular lattices like in Fig.
 \ref{lattices-e2}, i.e., with the Schl\"afli symbols $\{p,k\} =
 \{6,3\},\ \{4,4\},\ \{3,6\}$ can be embedded in $\mathbb{T}^2$,
$\mathbb{K}^2$, or $\mathbb{P}^2$. To do this we must solve the system of
equations
\begin{equation}
    V-E+F=\chi(M),~~pF=kV=2E,
    \label{eiler1}
\end{equation}
where $V,E,F$ are numbers of vertices, edges, and faces, respectively,
$\chi(M)$ is the Euler characteristic of a manifold $M$. Since
$\chi(\mathbb{T}^2)= \chi(\mathbb{K}^2)=0$ and $\chi(\mathbb{P}^2)=1$, we
see that the regular lattices are possible only in the torus and the Klein
bottle and impossible in the projective plane, as well as in any other
closed surfaces.
\par
Summarizing, for Euclidean metric there are only 3-valent hexagonal,
4-valent square and 6-valent triangular regular lattices in
$\mathbb{E}^2,$ $\mathbb{T}^2$, and $\mathbb{K}^2$.
\subsection{Hyperbolic Plane $\mathbb{H}^2$}
The hyperbolic (Lobachevsky) plane  $\mathbb{H}^2$ allows infinitely many
regular lattices. Poincar\'e proved that regular tilings $\{p,k\}$ of $\mathbb{H}^2$ exist for any $p, k \geq3$ satisfying
$\frac{1}{p}+\frac{1}{k}<\frac{1}{2}.$
\par
\begin{figure}
    \centering
\includegraphics[width=100pt,height=100pt]{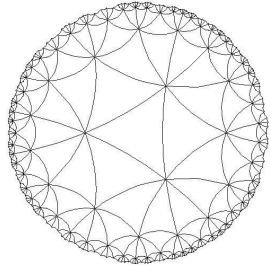}
\caption{Octivalent regular lattice $\{3,8\}$ in $\mathbb{H}^2$}
    \label{8-valent-in-h2}
\end{figure}

For example, let us consider the octivalent Moore
neighborhood used in the Life family and shown in the figure below.
\begin{center}
\setlength{\unitlength}{1.5pt}
\begin{picture}(30,30)(0,0)
\multiput(5,5)(10,0){3}{\circle*{2}}
\multiput(5,15)(10,0){3}{\circle*{2}}
\multiput(5,25)(10,0){3}{\circle*{2}}
\multiput(5,2)(10,0){3}{\line(0,1){26}}
\multiput(2,5)(0,10){3}{\line(1,0){26}}
\put(2,2){\line(1,1){26}}
\put(12,2){\line(1,1){16}}
\put(2,12){\line(1,1){16}}
\put(2,28){\line(1,-1){26}}
\put(12,28){\line(1,-1){16}}
\put(2,18){\line(1,-1){16}}
\end{picture}
\end{center}
The Moore neighborhood can not form regular
lattice in the Euclidean plane since the distances of surrounding cells
from the center are different. But in the hyperbolic plane regular
8-valent lattices exist and there are infinitely many of them. The simplest one is shown (using the
Poincar\'e disc model projection) in Fig. \ref{8-valent-in-h2}.
\subsection{Sphere $\mathbb{S}^2$}
\begin{figure}[!ht]
    \centering
$\overbrace{
\includegraphics[width=100pt,height=100pt]{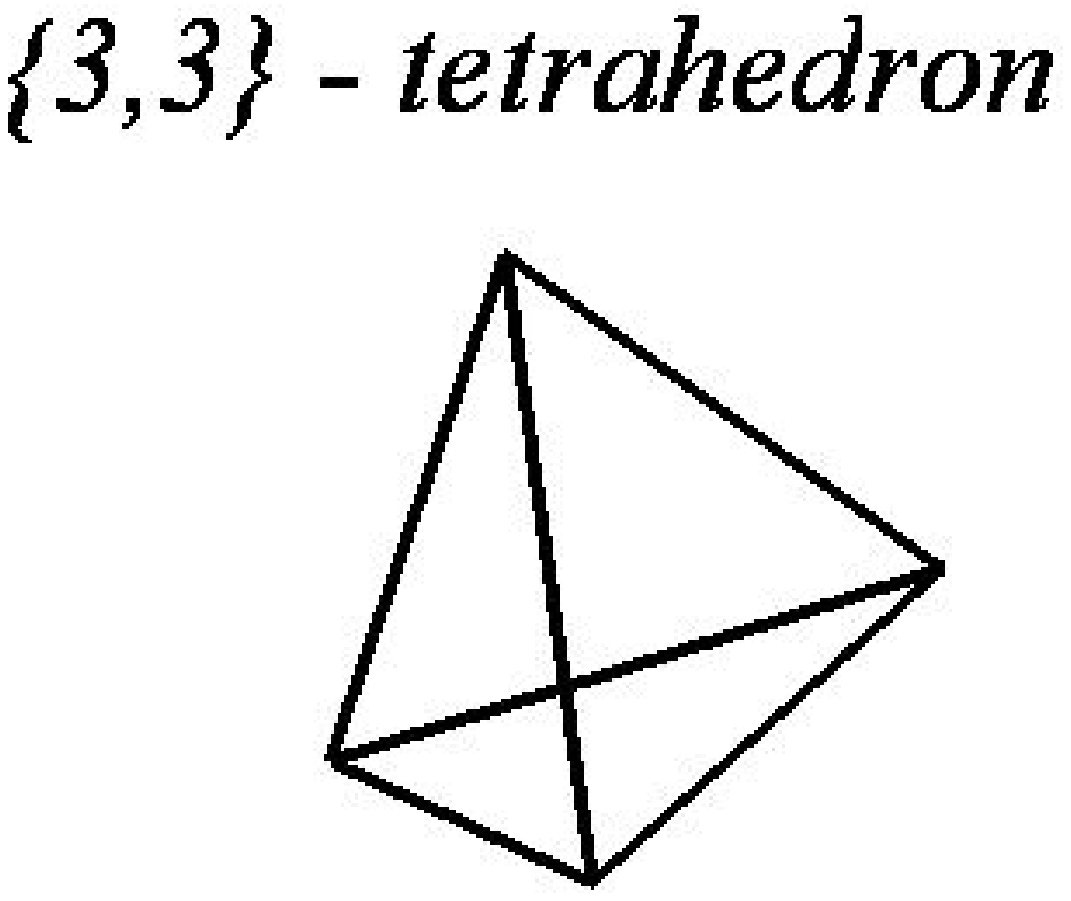}
\includegraphics*[width=100pt,height=100pt]{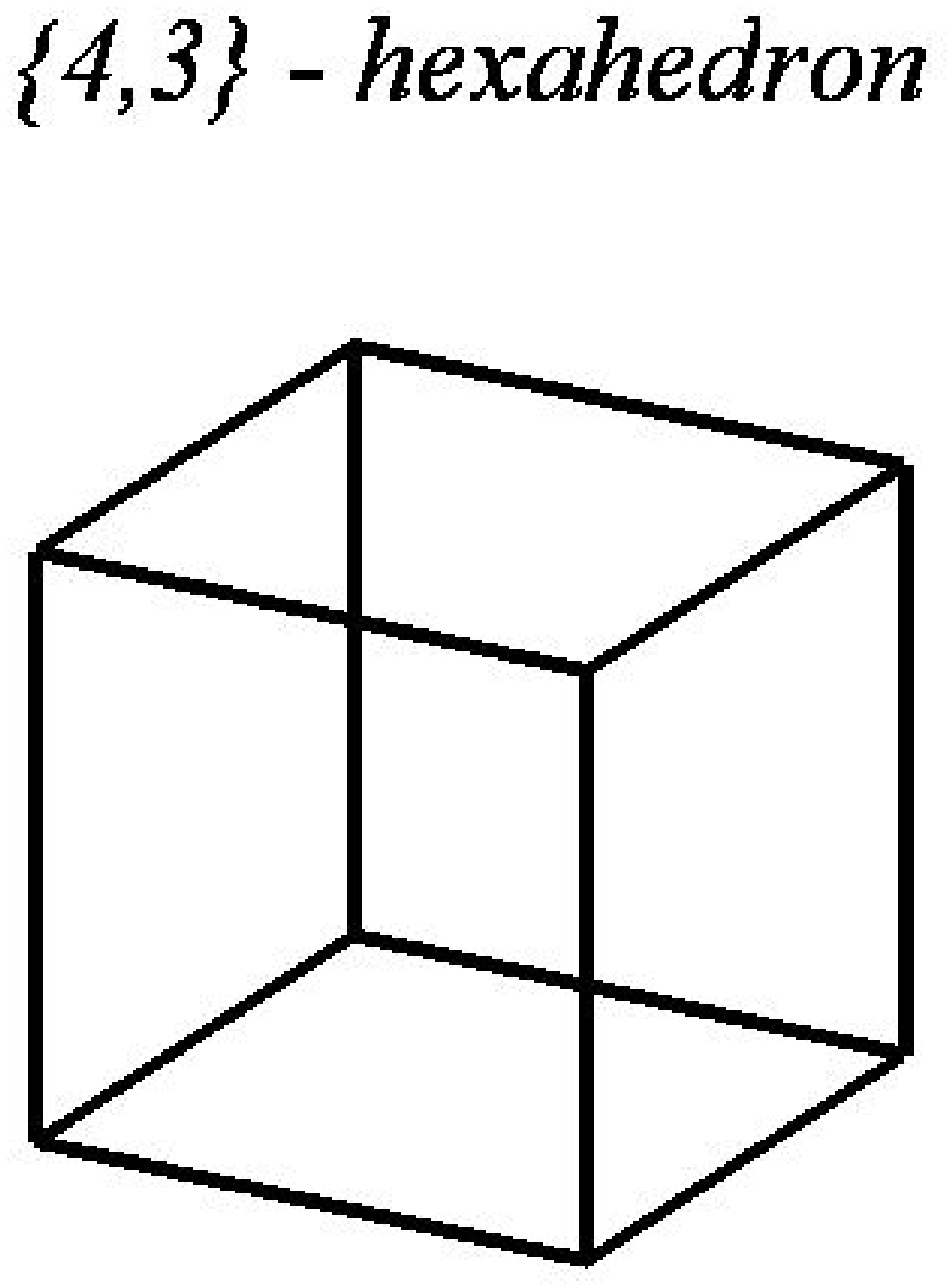}
\includegraphics*[width=100pt,height=100pt]{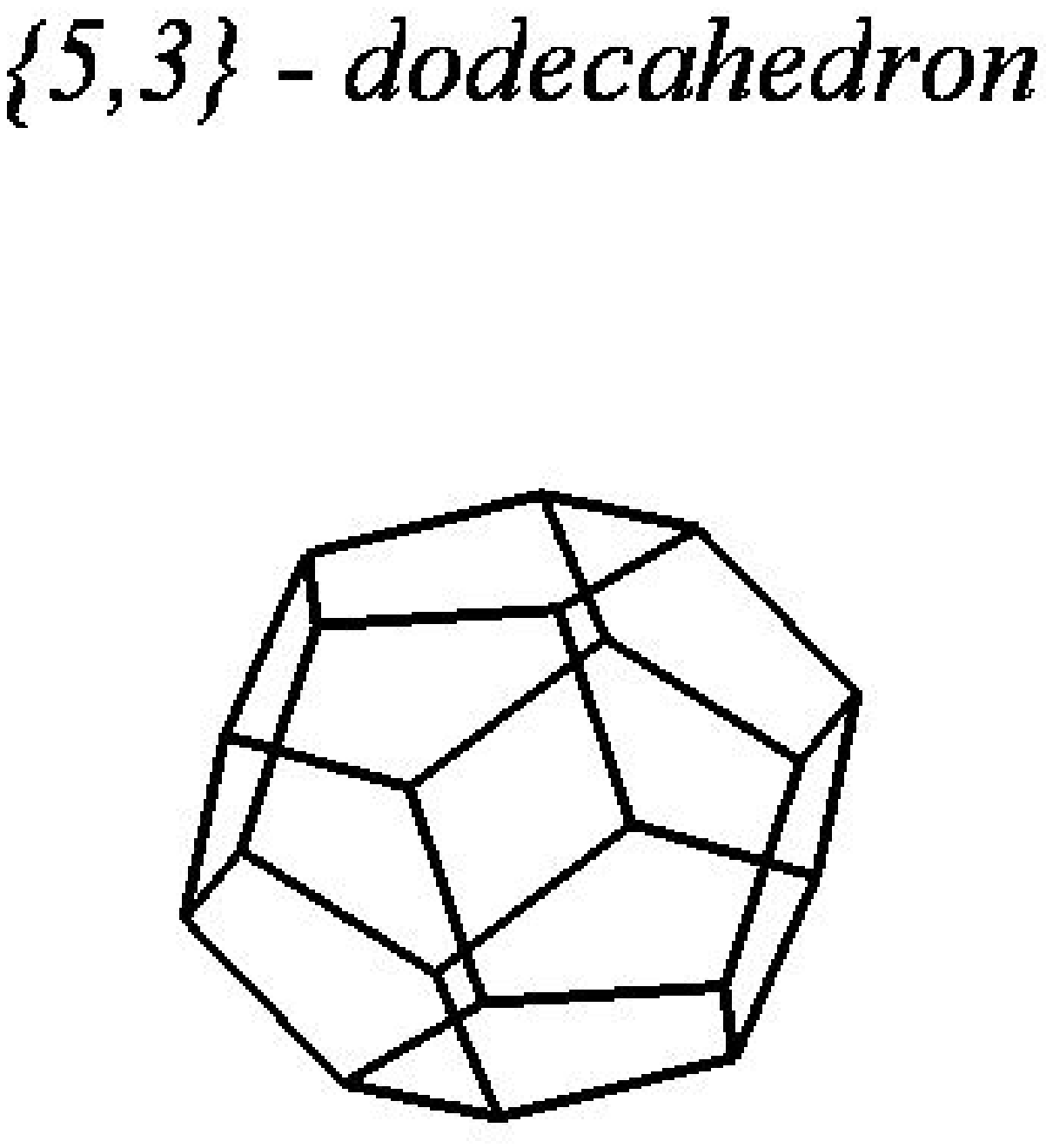}
}^{\text{\large\emph{\textbf{3-valent}}}
}$
    \par
$\overbrace{
\includegraphics[width=100pt,height=100pt]{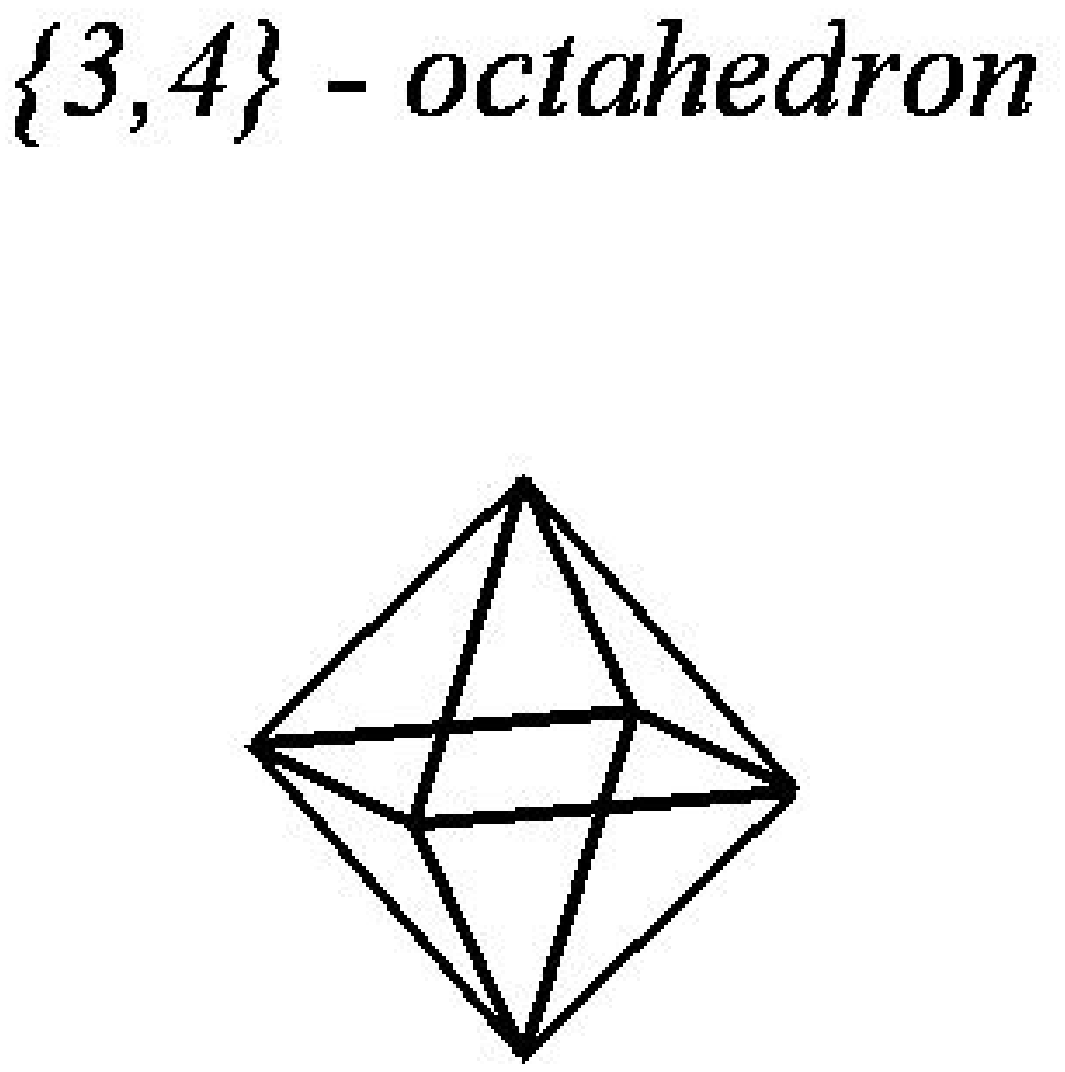}
}^{\text{\large\emph{\textbf{4-valent}}}
}
\hspace*{50pt}
\overbrace{
\includegraphics[width=100pt,height=100pt]{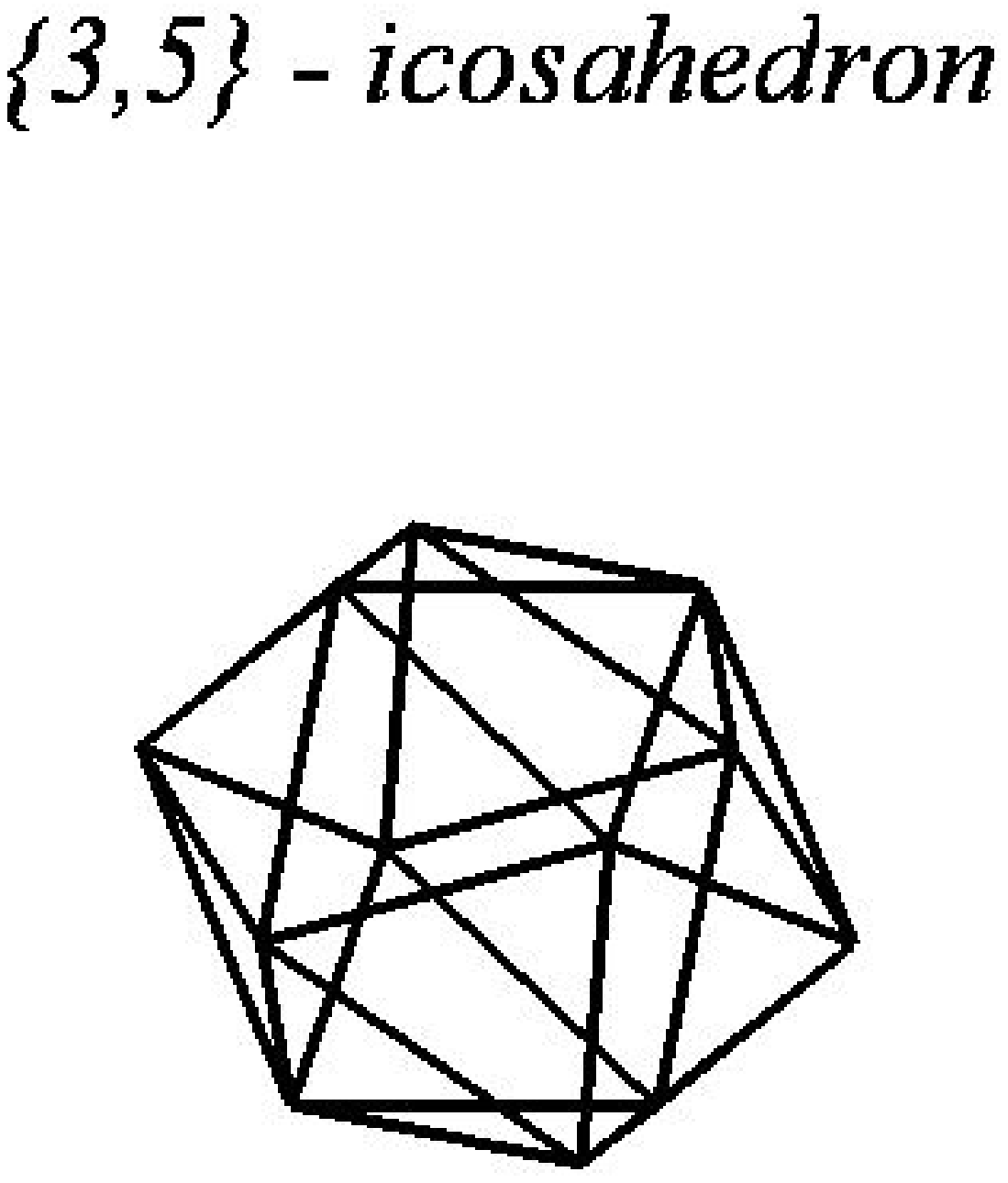}
}^{\text{\large\emph{\textbf{5-valent}}}
}$
\par
    \vspace*{-20pt}
    \caption{All regular lattices in $\mathbb{S}^2$}
    \label{all-in-s2}
\end{figure}

All regular lattices in the two-dimensional sphere correspond to the
 Platonic solids which are shown in Fig. \ref{all-in-s2}.
\par
In the sphere there are infinitely many other lattices, which are
close to regular and may serve as spaces for 3-valent symmetric automata. They are called \emph{fullerenes}.
\par
\begin{figure}[!htbp]
  \centering
\includegraphics[width=100pt,height=100pt]{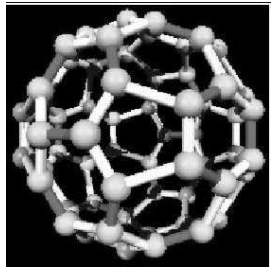}
    \caption{A model of $C_{60}$ carbon molecule (buckyball).}
    \label{buckyball}
\end{figure}

The fullerenes were first discovered in carbon chemistry in 1985, and this
discovery was rewarded with the 1996 Nobel Prize in Chemistry. A model of
the first revealed fullerene --- the carbon molecule $C_{60}$ --- is
displayed in Fig. \ref{buckyball} (the figure is borrowed from
\cite{Atiyah}).
Later there were discovered other forms of large carbon molecules with
structural properties of fullerenes (larger spherical fullerenes, carbon
nanotubes, graphenes). Their unique properties promise they will have an
important role in future technology, in particular, in nanotechnology
engineering.
\par
From a mathematical viewpoint, the structure of fullerene is a 3-valent
convex polyhedron with pentagonal and hexagonal faces. In terms of graphs,
fullerene can be defined as a 3-regular (3-valent) \emph{planar}, or
equivalently, embeddable in $M=\mathbb{S}^2$, graph with all faces of size
5 or 6.
\par
Let us generalize slightly this definition assuming that $M$ is not
necessarily $\mathbb{S}^2$, but may be closed surface of other type,
orientable or nonorientable. Then the Euler--Poincar\'e equation together
with the edge balance relations gives the system of equations
\begin{equation}
    V-E+f_5+f_6=\chi(M),~~3V=5f_5+6f_6=2E,
    \label{euler2}
\end{equation}
where $f_5$ and $f_6$ are numbers of pentagons and hexagons, respectively.
The general solution of this system is
\begin{eqnarray}
    f_5&=&6\chi(M),\label{f5}\\
    V&=&2f_6+10\chi(M),\\
    E&=&3f_6+15\chi(M).
\end{eqnarray}
We see that generalized fullerenes are possible only in the sphere
$\mathbb{S}^2$, in the projective plane $\mathbb{P}^2$, and in the torus
$\mathbb{T}^2$ and Klein bottle $\mathbb{K}^2$. Attempts to consider
surfaces with greater genus lead due to (\ref{f5}) to senseless negative
numbers of pentagons. Thus, we have for all generalized fullerenes:
$$
\begin{array}{llll}
V = 2f_6+20,& ~E = 3f_6+30,& ~f_5 = 12,& \text{~sphere~} \mathbb{S}^2;\\
V = 2f_6+10,& ~E = 3f_6+15,& ~f_5 = 6,& \text{~projective plane~}
\mathbb{P}^2;\\
V = 2f_6, & ~E = 3f_6,& ~f_5 = 0,&  \text{~torus~} \mathbb{T}^2,
\text{~Klein bottle~} \mathbb{K}^2.
\end{array}
$$
We see that any fullerene in $\mathbb{S}^2$ or in $\mathbb{P}^2$ contains
exactly 12 or 6 pentagons, respectively, and arbitrary number of hexagons.
In the case of torus or Klein bottle the fullerene structure degenerates
into purely hexagonal lattice without pentagons considered already in
subsection \ref{2D Euclidean Metric}. Note that in carbon chemistry
one-layer graphite sheets (and similar structures with few pentagons or
heptagons added) are called \emph{graphenes}.

\section{Canonical Decomposition of Some Rules from Life Family}

In this section we present the canonical decompositions of
relations for three rules from the Life family:
the standard \emph{\textbf{Conway's Life}},
\emph{\textbf{HighLife}} and  \emph{\textbf{Day\&Night}}.
\par
Recall that canonical decomposition \cite{Kornyak05} of relation on a set
of points is the representation of the relation as a combination of its
projections onto subsets of points. This decomposition is discrete analog
of compatibility analysis of algebraic and differential equations. To
apply our approach we interpret the local rule $x'_9=f\left(x_1,
x_2,\ldots,x_9\right)$ as a relation on 10 points $x_1, x_2,\ldots,x_9,
x'_9$.
\par
The \emph{\textbf{HighLife}} is interesting since for it the \emph{replicator} --- a self-reproducing pattern --- is known explicitly.
For \emph{\textbf{Conway's Life}}, the existence of replicators is proved,
but no example is known.
\par
The name \emph{\textbf{Day\&Night}} reflects the BW-symmetry of this rule.
Perhaps, it is the first (conceivably found ``by hand'') automaton from
the Life family with this symmetry property. Note that there are exactly
512 such automata in the Life family, and they all are generated easily
(for time $<$ 1 sec) by the C program mentioned in \cite{Kornyak05}.
\par
It might be easier to grasp the relations, if they are written in the form
of polynomials over the field $\F_2$. In the below formulas we use
\emph{elementary symmetric polynomials} in variables $x_1,\ldots,x_8.$
Recall that
the $k$th (of degree $k$) elementary symmetric polynomial of
$n$ variables
$x_1,\ldots,x_{n}$ is defined by the formula:
$$
\esymm{k}{x_1,\ldots,x_{n}} =
\sum\limits_{1\leq i_1<i_2<\cdots<i_{k}\leq n}x_{i_1}x_{i_2}\cdots x_{i_{k}}.
$$
Hereafter we use the  notations
$
\sigma_k\equiv\esymm{k}{x_1,\ldots,x_8}\!,$ $\sigma^i_k\equiv\esymm{k}{x_1,\ldots,
\widehat{x_i},\ldots,x_8},$
$\sigma^{ij}_k\equiv\esymm{k}{x_1,\ldots,
\widehat{x_i},\ldots,\widehat{x_j},\ldots,x_8}.
$ The indices $i,j,k,l$ satisfy $1\leq i<j<k<l\leq 8.$
\par
The polynomials representing \emph{\textbf{Conway's Life}}, \emph{\textbf{HighLife}} and \emph{\textbf{Day\&Night}}
 take the forms, respectively
\begin{eqnarray}
    P_{\textbf{Conway's Life}}&=& x'_9
    +x_9
    \left\{
    \sigma_7
    +\sigma_6
    +\sigma_3
    +\sigma_2
    \right\}
    +\sigma_7
    +\sigma_3,
    \label{polyCL}\\
    P_{\textbf{HighLife}}&=& x'_9
    +x_9
    \left\{
    \sigma_3
    +\sigma_2
    \right\}
    +\sigma_6
    +\sigma_3,
    \label{polyHL}\\
    P_{\textbf{Day\&Night}}&=& x'_9
    +x_9
    \left\{
    \sigma_7
    +\sigma_6
    +\sigma_5
    +\sigma_4
    \right\}
    +\sigma_8
    +\sigma_7
    +\sigma_6
    +\sigma_3.
    \label{polyDN}
\end{eqnarray}
Note that these
polynomials
have degrees 8, 6, 8 and numbers of terms 185, 169, 256,
respectively, so application of the Gr\"obner basis technique to their
analysis may take some time (about 1 hour on 1.8GHz AMD
Athlon notebook with the Maple 9 Gr\"obner procedure). Our program computes
the decompositions for time $<$ 1 sec.
\par
The decompositions are:
\begin{itemize}
    \item
\emph{\textbf{Conway's Life}}:
\begin{eqnarray}
x'_9
\left\{
\sigma_3
+\sigma_2
+1
\right\}
+\sigma_7
+\sigma_3&=&0,
\label{poly2red}
\\
x'_9x_9
\left\{
\sigma^i_2
+\sigma^i_1
\right\}
+x'_9
\left\{
\sigma^i_2
+1
\right\}
+x_9
\left\{
\sigma^i_7
+\sigma^i_6
+\sigma^i_3
+\sigma^i_2
\right\}&=&0,
\label{poly1red}
\\
x'_9
\left\{
\sigma^{i}_3
+\sigma^{i}_2
+\sigma^{i}_1
+1
\right\}&=&0,
\label{poly12red}
\\
x'_9\left(x_9
+1\right)
\left\{
\sigma^{ij}_3
+\sigma^{ij}_2
+\sigma^{ij}_1
+1
\right\}&=&0,
\label{poly11red}
\\
x'_9x_{i}x_{j}x_{k}x_{l}&=&0.
\label{poly0123red}
\end{eqnarray}
    \item \emph{\textbf{HighLife}}
\begin{eqnarray}
x'_9
\left\{
\sigma_3+\sigma_2+1
\right\}
+\sigma_7+\sigma_3
&=&0,
\\
x'_9x_9
\left\{
\sigma^i_2+\sigma^i_1
\right\}
+x'_9
\left\{
\sigma^i_5+\sigma^i_2+1
\right\}
+x_9
\left\{
\sigma^i_7+\sigma^i_6+\sigma^i_3+\sigma^i_2
\right\}
&=&0,
\\
x'_9
\left\{
\sigma^{i}_7
+\sigma^{i}_3
+\sigma^{i}_2
+\sigma^{i}_1
+1
\right\}&=&0,
\label{poly12redHL}
\\
x'_9x_9
\left\{
\sigma^{ij}_3+\sigma^{ij}_2+\sigma^{ij}_1+1
\right\}\hspace*{150pt}&&\nonumber \\
+x'_9
\left\{
\sigma^{ij}_6+\sigma^{ij}_5+\sigma^{ij}_4+\sigma^{ij}_3
+\sigma^{ij}_2+\sigma^{ij}_1+1
\right\}
&=&0,\\
x'_9x_9x_i x_j x_k x_l&=&0.
\end{eqnarray}
    \item \emph{\textbf{Day\&Night}}
\begin{eqnarray}
x'_9
\left\{
\sigma_7
+\sigma_6+\sigma_5+\sigma_4
+1
\right\}
+\sigma_8+\sigma_7
+\sigma_6+\sigma_3&=&0,
\label{poly2redDN}
\\
x'_9x_9
\left\{
\sigma^i_6+\sigma^i_5+\sigma^i_4+\sigma^i_3
\right\}
+
x'_9
\left\{
\sigma^i_7+\sigma^i_6+\sigma^i_5+\sigma^i_2+1
\right\}
&& \nonumber\\
+
x_9
\left\{
\sigma^i_7+\sigma^i_3
\right\}
+\sigma^i_6&=&0,\\
x'_9
\left\{
\sigma^{ij}_5+\sigma^{ij}_4+\sigma^{ij}_3+\sigma^{ij}_2+\sigma^{ij}_1+
1
\right\}
+\sigma^{ij}_6&=&0.
\label{ijDN}
\end{eqnarray}
Note that system (\ref{ijDN}) of prime relations can be combined into the
reducible relation (for terminology like \emph{prime} and \emph{reducible}
see \cite{Kornyak05})
$$
x'_9
\left\{
\sigma^i_4+\sigma^i_2+1
\right\}
+\sigma^{i}_6=0,
$$
which looks nicer (the polynomial representation of relations is somewhat
artificial), but depends on larger set of variables.
\end{itemize}
We see that the decomposition for \emph{\textbf{Day\&Night}} differs
essentially from the decompositions for \emph{\textbf{Conway's Life}} and
\emph{\textbf{HighLife}} having resembling structures.

\section{Conclusions}
We proved that the cellular automata from the Life family are nothing but
binary automata with the local rules symmetric with respect to all
permutations of the outer cells in the neighborhood.
\par
Then we showed that
the number of non-equivalent with respect to renaming of states $k$-valent
symmetric binary local rules is equal to $2^{2k+1}+2^k$. Considering the
3-valent case --- the next step up after the 2-valent Wolfram's elementary automata
--- we see that the total number of non-equivalent symmetric rules is
only 136.
\par
All interesting 3-valent 2-dimensional regular (or almost regular)
 lattices are:
\begin{itemize}
    \item hexagonal lattice $\{6,3\}$ in the plane $\mathbb{E}^2$, in the
     torus $\mathbb{T}^2$ and in the Klein bottle $\mathbb{K}^2$
    \item tetrahedron  $\{3,3\}$, hexahedron (cube) $\{4,3\}$ and
    dodecahedron $\{5,3\}$ in the sphere $\mathbb{S}^2$
    \item fullerenes in $\mathbb{S}^2$  and in the projective plane
     $\mathbb{P}^2$.
\end{itemize}
Combining these lattices with 136 symmetric rules we obtain a class of
cellular automata which looks like quite available for systematic study.
Since these automata have more interesting geometry than the elementary
ones we may expect more interesting behavior of them.
\section*{Acknowledgments}
I am very grateful to Vladimir Gerdt for detailed discussions about the
present paper.
This work was supported in part by the
grants
04-01-00784
from the Russian Foundation for Basic Research
 and
2339.2003.2
 from the Russian Ministry of Industry, Science, and
Technologies.

\end{document}